\documentclass[aps,prl,twocolumn,groupedaddress]{revtex4}
\usepackage{amssymb}
\usepackage{amsmath}
\usepackage{amsthm}
\usepackage{amsfonts}
\usepackage{upgreek}
\usepackage{enumerate}

\usepackage{graphicx}
\usepackage{dcolumn}
\usepackage{bm}

\begin{document}

\title{Continuous Lyman-alpha generation by four-wave mixing in mercury for 
       laser-cooling of antihydrogen}


\author{D. Kolbe}
\email[]{kolbed@uni-mainz.de}
\author{A. Beczkowiak}
\author{T. Diehl}
\author{A. Koglbauer}
\author{A. M{\"{u}}llers}
\author{M. Scheid}
\author{M. Stappel}
\author{R. Steinborn}
\author{J. Walz}

\affiliation{%
Institut f{\"{u}}r Physik, Johannes 
           Gutenberg-Universit{\"{a}}t Mainz and Helmholtz Institute Mainz, D-55099 Mainz
}%


\date{\today}

\begin{abstract}

Cooling antihydrogen atoms is important for future experiments both to
test the fundamental CPT symmetry by high-resolution laser
spectroscopy and also to measure the gravitational acceleration of
antimatter.  Laser-cooling of antihydrogen can be done on the strong
1\,S--2\,P transition at the wavelength of Lyman-alpha (121.6\,nm). 
A continuous-wave laser at the Lyman-alpha wavelength based on solid-state fundamental lasers is described.
By using a two-photon and a near one photon resonance a scan across the whole phasematching curve of the four-wave mixing process is possible. Furthermore the influence of the beam profile of one fundamental beam on the four-wave mixing process is studied. 

\end{abstract}


\maketitle

\section{Introduction}
Future high-resolution laser-spectroscopy of antihydrogen in a
magnetic trap can provide very stringent tests of the fundamental
symmetry between matter and antimatter (CPT symmetry)
\cite{Bluhm1999}.  The two-photon transition from the 1\,S ground
state to the 2\,S metastable excited state is a good candidate
for such precision experiments because it has a natural linewidth of
only 1.6\,Hz at a transition frequency of 2\,466\,THz.  The absolute
frequency of this transition has already been measured in ordinary
hydrogen with enormous precision \cite{Niering2000,Fischer2004}.
However, the 1S--2S ($F=1, m_F= 1 \rightarrow F=1, m_F=1$) transition
frequency has a residual dependence on the magnetic field of 186\,kHz
per Tesla.  This will broaden and shift the spectral line of
antihydrogen atoms in a magnetic trap
\cite{Cesar2001}.  It will thus be very
important to cool antihydrogen atoms, thereby reducing their spatial
spread in the inhomogeneous magnetic field of the trap.
Laser cooling of ordinary hydrogen atoms in a magnetic trap to the
milli-Kelvin temperature range has been demonstrated
\cite{Setija1993}.  A pulsed laser source has been used to drive the
strong Lyman-alpha transition at 121.6\,nm wavelength from the 1S
ground state to the 2P excited state.

In addition to testing CPT there is also the intriguing prospect to
measure the gravitational acceleration of antimatter for the first
time using antihydrogen atoms \cite{Gabrielse1988}. The thermal
motion of antihydrogen is a critical factor in this type of
experiments and laser-cooling at Lyman-alpha to milli-Kelvin
temperatures will be very beneficial.  Ultimately, ultracold temperatures in
the sub-milli-Kelvin range are desirable for practical experiments.
These temperatures are beyond standard laser-cooling limits for \mbox{(anti-)}
hydrogen.  Novel cooling schemes for ultracold temperatures
have been proposed \cite{Walz2004,Kellerbauer2006}.

Producing coherent radiation at 121.6\,nm (Lyman-alpha) is a challenge
as there are no tunable lasers and nonlinear frequency-doubling
crystals available for that spectral region.  Sum-frequency generation
of several incident laser beams utilizing the nonlinear susceptibility
of atomic vapors and gases is commonly used to produce coherent
radiation in the VUV.  Four-wave sum-frequency mixing produces the
sum-frequency of three fundamental colors \cite{Vidal1992} and has been
employed to generate {\em{pulsed}} laser radiation at Lyman-alpha,
typically using Krypton gas
\cite{Mahon1978,Cotter1979,Wallenstein1980,Cabaret1987,Marangos1990}.

{\em{Continuous}} coherent radiation at Lyman-alpha can have distinct
advantages for laser-cooling of antihydrogen. For example, the cooling rate is not limited by the pulse to pause ratio and the smaller linewidth reduces pumping in untrapped states. An important
difference, however, is that the power levels of continuous
fundamental beams are many orders of magnitude lower than the peak
powers typically used in pulsed Lyman-alpha generation.  Continuous
Lyman-alpha generation therefore uses resonances and near-resonances
in the nonlinear optical medium.  

In this paper a continuous coherent Lyman-alpha source is described 
which uses a four-wave mixing process in mercury with an exact two-photon resonance
and a near one photon resonance of the fundamental beams. The paper is organized as follows: First the lasersystem for the fundamental beams and the Lyman-alpha production is explained. In the next part the influence of the UV beam profile on the four-wave mixing process is discussed and in the last part phasematching of the four-wave-mixing process at low mercury temperature is shown.

\section{Lasersetup and Lyman-alpha production}

\begin{figure}
  \includegraphics[width=4.5cm]{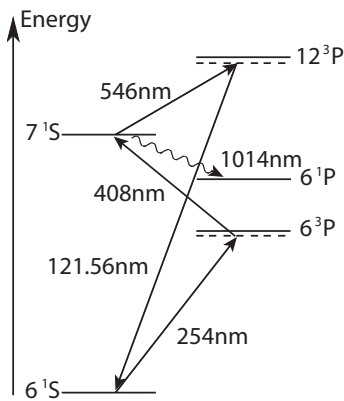}
  \caption{Four-wave mixing scheme to produce continuous laser
           radiation at Lyman-alpha in mercury vapor. The UV laser at 254\,nm wavelength is tuned closed to the 6$^1$S--6$^3$P transition. The blue laser at 408\,nm establishes the two-photon resonance to the 6$^1$S--7$^1$S transition. Taken from \cite{Scheid2009}.}
 \label{kolbe_fig1}
\end{figure}

Figure \ref{kolbe_fig1} shows the level scheme of mercury used in the four-wave mixing process. A UV beam at 254\,nm and a blue beam at 408\,nm wavelength establish the two photon resonance between the 6$^1$S ground state and the 7$^1$S state of mercury. The third beam at 545\,nm is fixed so that the sum frequency of the three fundamental beams is at the Lyman-alpha wavelength. The infrared fluorescence light from the decay of the 7$^1$S state to the 6$^1$P state provides a diagnostic tool for the two-photon-condition \cite{Beyer2009}.

\begin{figure}
  \includegraphics[width=8.0cm]{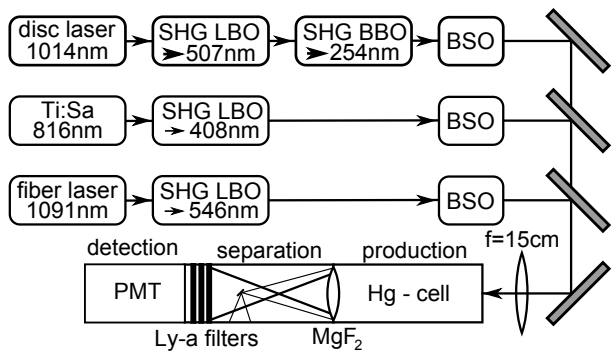}
  \caption{Setup of the continuous Lyman-alpha source. Upper part: Laser setup of the fundamental beams. Lower part: Mercury cell. The overlapped beams are focused in the mercury cell were the four-wave-mixing process takes place. A MgF$_2$ lense separates the production region from the separation and detection region. BSO: beam shaping optics, BBO/LBO: $\beta$-barium borate/lithium triborate frequency doubling crystals.}
 \label{kolbe_fig2}
\end{figure}

A schematic of the experimental setup for the Lyman-alpha production is shown in Fig.~\ref{kolbe_fig2}. The laser system generating the three fundamental beam is in the upper part of the figure. The beam at $254\,\text{nm}$ is provided by a frequency-quadrupled Yb:YAG disc laser. Frequency-quadrupling is done with two successive frequency-doubling cavities, the first one using a lithium triborate crystal (LBO) as nonlinear medium, the second one using a $\beta$-barium borate crystal (BBO). Details of this system have been described elsewhere \cite{Scheid2007}. From $2\,\text{W}$ of infrared light at $1015\,\text{nm}$ we get up to $200\,\text{mW}$ of UV radiation. This system is in principle capable to produce up to 750\,mW of UV light. To prevent the BBO crystal from damage caused by the high green powers we work at reduced powers of up to 300\,mW in the UV. The second fundamental beam at $408\,\text{nm}$ is produced by a frequency-doubled titanium:sapphire laser. The external frequency-doubling cavity uses LBO as nonlinear medium. From $1.5\,\text{W}$ of IR light at $816\,\text{nm}$ we get up to $500\,\text{mW}$ of blue light. The typical day to day power of this laser system is 300\,mW. The third fundamental beam at $545\,\text{nm}$ is produced with a 10\,W fiber laser system at 1091\,nm and a modified commercial frequency-doubling cavity. This system is capable of producing up to $4\,\text{W}$ of green light \cite{Markert2007}. However, at these high powers amplified back-reflections tend to damage the entrance facet of the amplification fiber. For the present experiments we therefore operate the fiber laser at $740\,\text{mW}$, a very conservative rating, which still gives $280\,\text{mW}$ of green light. 

The astigmatism of each of the fundamental beams is compensated with a pair of cylindrical lenses. All three beam divergences are matched by telescopes and the beam diameters are expanded to allow tighter focusing. The beams are then overlapped with dichroic mirrors and focused into the mercury cell using a fused silica lens with a focal length of $15\,\text{cm}$ (Figure \ref{kolbe_fig2} lower part). The average confocal parameter of the fundamental beams is  b=1.6\,mm. Details of the mercury cell are described  elsewhere \cite{Eikema2001}. The alignement of the fundamental beams is very critical. The overlap of the UV and blue foci can be improved by maximizing the IR fluorescence at $1014\,\text{nm}$. The overlap of the green beam can then be adjusted using the Lyman-alpha signal. The Lyman-alpha beam is separated from the fundamental beams using the dispersion of a MgF$_2$ lens ($f=21.5\,\text{cm}$ at $546\,\text{nm}$,$f=13\,\text{cm}$ at Lyman-alpha). A tiny mirror is placed in the focus of the fundamental beams to reflect them to the side. Hence, the mirror just casts a shadow in the Lyman-alpha beam, causing $\approx 30\%$ loss. Light from the fundamental beams is suppressed further by three VUV interference filters (Acton, 122-N and 122-XN) and the radiation at Lyman-alpha is then detected with a solar-blind photomultiplier (Hamamatsu, R6835). A residual background from the UV beam is eliminated by chopping the green laser at 1\,Hz and subtracting the background from the Lyman-alpha signal. We get 8000 Lyman-alpha counts/s with laser powers in the mercury cell of 180\,mW at $254\,\text{nm}$, 200\,mW at $408\,\text{nm}$ and 260\,mW at $546\,\text{nm}$ and with a detuning to the $6^3$P level of the $^{202}\text{Hg}$ isotope of -400\,GHz \cite{Scheid2009}. The overall detection efficiency of Lyman-alpha due to the absorption in the MgF$_2$ lens, the small mirror, the filters and the photomultiplier efficiency is $3*10^{-5}$. Therefore the Lyman-alpha power generated is about 0.4 nW. We observed no saturation of the VUV yield which is proportional to the power-product of the fundamental beams. The Lyman-alpha generated can thus readily be enhanced by increasing the power in the fundamental beams. The power levels required for Doppler-cooling of antihydrogen in a magnetic trap are rather low: With $1\,\text{nW}$ of Lyman-alpha we expect cooling times in the order of minutes \cite{Walz2001}.

\section{Influence of the UV beam profile on the four-wave-mixing}

\begin{figure}[htb]
\centerline{\includegraphics[width=9cm]{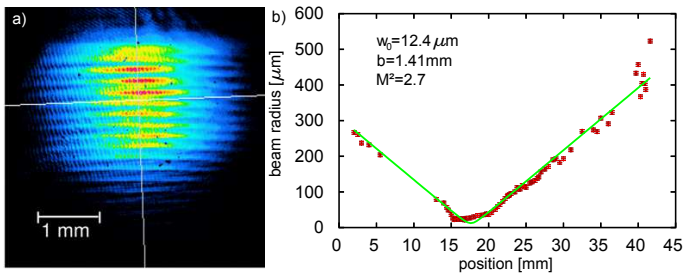}}
\caption[Ly-a measurement]{(a) Beam profile of the collimated UV beam. (b) Beam propagation of the UV beam near the focus.} 
\label{Fig:3}
\end{figure}

Figure~\ref{Fig:3}a shows the beam profile of the collimated UV beam. The large walk-off angle of the second frequency-doubling crystal results in an extrem non-gaussian beam profile. The beam propagation near the focus (see Figure \ref{Fig:3}b) yields a beam parameter product of $M^2=2.7$ along the disturbed axis of the beam. Along the non-disturbed axis the beam parameter product is $M^2=1.4$. To measure the influence of such a beam profile on the four-wave-mixing process a spatial filter was installed in the UV beam. It consists of two lenses building up a telescope with a pinhole in the focus of the UV beam. Such a configuration reduces higher modes in a non-gaussian beam which results in a reduced beam parameter product. With a transmission of $\approx 80\%$ through the spatial filter a reduction in the beam parameter product to $M^2=1.4$ could be achieved. To directly compare the four-wave-mixing with different beam-profiles the pinhole of the spatial filter was removed during a measurement. The 20$\%$ higher beam power due to the absence of the pinhole does not result in an increase of the Lyman-alpha signal. Hence, the higher non-Gaussian modes in the beam do not participate in the four-wave-mixing process. Therefore, we can conclude that the power which effectively contributes in the four-wave mixing is just the fraction embedded in the Gaussian mode of the fundamental beams.

\section{Phasematching}

\begin{figure}[htb]
\centerline{\includegraphics[width=6cm,angle=-90]{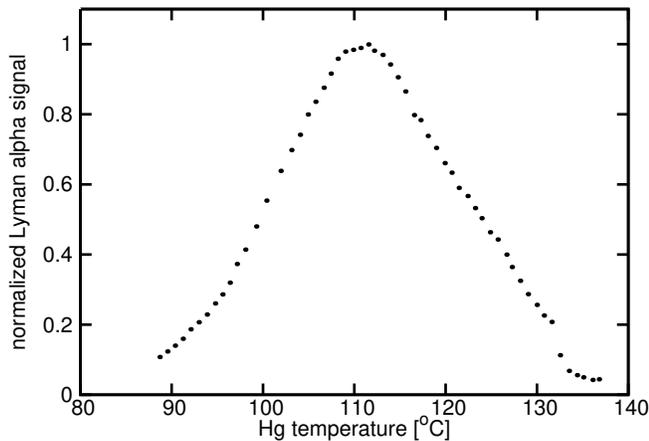}}
\caption[Temperature-scan]{Normalized Lyman-alpha yield as a function of the temperature of the mercury cell.}
\label{Fig:4}
\end{figure}

Figure~\ref{Fig:4} shows the Lyman-alpha signal as a function of the temperature of the mercury cell. The UV laser detuning to the $6^3$P level was chosen to \mbox{-50\,GHz} and the blue laser frequency is tuned to match the two-photon resonance of the $^{202}$Hg isotope. Optimal phasematching for sum-frequency four-wave-mixing occurs at $b\Delta k=-4$ \cite{Bjorklund1975}, where $b$ is the confocal parameter and $\Delta k =k_{Ly-\alpha}-k_{254nm}-k_{408nm}-k_{546nm}$ the wavevector mismatch. The wavevector mismatch $\Delta k$ is proportional to the mercury density $N$ and can be changed with the mercury temperature. Due to the near $6^1\text{S} - 6^3\text{P}$ resonance the main contribution to $\Delta k$ is the dispersion of the UV beam. A smaller detuning to the $6^1\text{S} - 6^3\text{P}$ resonance, compared to earlier experiments \cite{Eikema2001}, leads to an increased dispersion of the UV beam and the optimal phase matching is achieved at lower mercury temperatures. This enables scanning the phase matching curve of this four-wave mixing process without exceeding the maximum temperature of the cell. The maximum of the Lyman-alpha yield at a UV detuning of -50\,GHz is at about 110$^\circ$C, which corresponds to a mercury density of  $1.1 \times 10^{22}\,\textrm{m}^{-3}$. The data points were normalized to the powers of the fundamental beams and to the peak maximum to cancel out power drift in the fundamental beams. Calculations of phasematching temperature as performed in \cite{Pahl2005} still differ in temperature compared to the experimental data. In Table \ref{table} a comparison of cw Lyman-alpha sources and their efficiencies, powers and phasematching temperatures is shown. The increased efficiency at lower phasematching temperature originates from reduced Doppler-broadening.  

\begin{table}
\begin{tabular}{c|c|c|c}
Ref. & efficiency & power  & Hg temperature \\ 
 & [nW/W$^3$]& [nW]  & [$^\circ$C]\\
\hline
\cite{Eikema1999}& 3.5 & 0.5 & 220\\
\cite{Eikema2001}& 58.5 & 20 & 220\\
\cite{Scheid2009}& 55.6 & 0.4 & 200\\
this work & 69.0 & 0.07 & 110
\end{tabular}
\caption{Comparison of four-wave mixing efficiency, power and used mercury temperature of cw Lyman-alpha sources.}
\label{table}
\end{table}

With the same spatial filtering technique described earlier we checked the influence of the UV beam profile on the phasematching. Perfect phasematching of the filtered and non-filterd fundamental beam occures at the same mercury temperature. 

\section{Conclusion}

In conclusion, a continuous-wave Lyman-alpha source is demonstrated by four-wave mixing of three solid-state fundamental beams in mercury. A small detuning of one fundamental laser to a resonance of mercury enables scanning across the whole phasematching curve. It is shown that only the embedded Gaussian mode of the fundamental beams participates in the four-wave mixing process.

\begin{acknowledgments}

We gratefully acknowledge support by the DFG and the BMBF.
\end{acknowledgments}


\end{document}